\ifx\pdfminorversion\undefined\else\pdfminorversion7\fi
\ifx\pdfsuppresswarningpagegroup\undefined\else\pdfsuppresswarningpagegroup1\fi
\PassOptionsToPackage{dvipsnames}{xcolor}
\PassOptionsToPackage{unicode}{hyperref}
\RequirePackage{textgreek}
\documentclass[a4paper,11pt]{article}
\usepackage{pos}
\usepackage{graphicx}
\usepackage{amsfonts} 
\usepackage{amsmath} 
\usepackage{caption}
\usepackage{slashed}

\usepackage[dvipsnames]{xcolor}

\renewcommand\color[1]{\relax}
\usepackage{pagecolor}

\graphicspath{{figs/}}

\title{Calculation of neutron electric dipole moment due to the QCD topological term, Weinberg three-gluon operator and the quark chromoelectric moment}

\ShortTitle{Calculation of nEDM on the lattice}

\author*[a]{Tanmoy Bhattacharya}
\author[a]{Vincenzo Cirigliano}
\author[a]{Rajan Gupta}
\author[a]{Emanuele Mereghetti}
\author[b]{Boram Yoon}
\affiliation{T-2, Theoretical Division, Los Alamos National Laboratory,\\
          MS B285, Los Alamos, NM 87545, USA}
\affiliation{CCS-7, Computer, Computational, and Statistical Sciences, Los Alamos National Laboratory,\\
          MS B287, Los Alamos, NM 87545, USA}
\emailAdd{tanmoy@lanl.gov}
\emailAdd{cirigliano@lanl.gov}
\emailAdd{rg@lanl.gov}
\emailAdd{emereghetti@lanl.gov}
\emailAdd{boram@lanl.gov}

\abstract{
  We present results for the neutron electric dipole moment due to the dimension 4 and dimension 6 gluonic CP violation, and the isovector quark chromoelectric dipole moment using clover valence quarks on HISQ dynamical ensembles generated by the MILC Collaboration. For the gluonic operators, we use the gradient flow scheme to obtain divergence-free continuum results. For the chromoelectric dipole moment operator, we use the unflowed local operator but discuss how the quadratically divergent mixing with the pseudoscalar operator can be controlled nonperturbatively. 
}

\FullConference{%
 The 38th International Symposium on Lattice Field Theory, LATTICE2021
  26th-30th July, 2021
  Zoom/Gather@Massachusetts Institute of Technology
}

\begin{document}
\maketitle
\section{Introduction}

The standard model of particle physics has a tiny CP violation (CPV) in the weak sector arising from a phase in the Cabibbo-Kobayashi-Maskawa~\cite{PhysRevLett.10.531,10.1143/PTP.49.652} mixing matrix of the quarks, and possibly a similar phase in the  Pontecorvo-Maki-Nakagawa-Sakata~\cite{10.1143/PTP.28.870,Pontecorvo:1957qd} matrix in the neutrino sector.  The possible CPV in the strong sector due to topological effects (the $Theta$-term) is known to be anomalously small~\cite{Baker:2006ts,Graner:2016ses}.  On the other hand, the observed baryon asymmetry of the universe~\cite{Bennett:2003bz} is difficult to explain without additional CPV processes~\cite{Sakharov:1967dj,Dolgov:1991fr,Coppi:2004za}. CPV is, therefore, a promising signature for constraining models of physics beyond the standard model (BSM).

To analyze the effect of CPV in a model-independent way, we start by classifying the CPV operators by their mass dimension.  At dimension 3, we can have CPV quark masses \(m_5 \bar\psi \gamma_5 \psi\).  These are necessarily flavor-singlet, since flavored axial rotations can be used to remove the rest.  The flavor-singlet axial symmetry is, however, anomalous---it rotates the CPV mass term into the topological charge operator \(G_{\mu\nu}\tilde G^{\mu\nu}\).  Thus up to dimension-4, there is only one CPV operator that we can take to be either a singlet CPV quark mass, or the gluonic topological term. 

At dimension 5, we have two kinds of operators that are suppressed by \(\Lambda_{\rm QCD}v_{\rm EW}/M^2_{\rm BSM}\), where \(\Lambda_{\rm QCD}\sim 300\rm MeV\) is a typical hadronic scale, \(v_{\rm EW} \sim 100\rm GeV\) is the electroweak scale, and \(M_{\rm BSM}\sim 1\rm TeV\) is the expected BSM scale: these operators are the electric dipole moment \(\bar\psi\Sigma_{\mu\nu}\tilde F^{\mu\nu}\psi\) and the chromoelectric dipole moment \(\bar\psi\Sigma_{\mu\nu}\tilde G^{\mu\nu}\psi\) of each quark. At dimension 6, we have a number of terms, all suppressed by \(\Lambda_{\rm QCD}^2/M^2_{\rm BSM}\): these are the gluon chromoelectric moment given by the Weinberg CPV three-gluon operator \(G_{\mu\nu}G_{\lambda\nu}\tilde G_{\mu\lambda}\), and various CPV four-Fermi operators.
At or below the hadronic scale, the leading effects of the high-scale CPV manifest as electric dipole moments of elementary particles, and as CPV interactions, for example, between pions and nucleons~\cite{Pospelov:2005pr}. Here we discuss the calculation of the electric dipole moments of the nucleons on the lattice.


The electric dipole moment of nucleons can be obtained from the electromagnetic vector form factors.  By Lorentz symmetry, there are four of these: the Dirac \(F_1\), the Pauli \(F_2\), the electric dipole \(F_3\), and the anapole \(F_A\) form factors.  These are related to the Sachs electric \(G_E \equiv F_1 - (q^2/4M_N^2) F_2\) and magnetic \(G_M \equiv F_1 + F_2\) form factors, which have intuitive interpretations in the Breit frame.  The form factor \(F_3\) breaks CP symmetry, whereas \(F_A\) breaks PT. All the form factors can be obtained by decomposing the matrix elements of the vector current in the nucleon state into its Lorentz covariant pieces:
\begin{eqnarray}
\langle N | V_\mu(q) | N \rangle & = &
   \overline {u}_N \left[ 
         \gamma_\mu\;F_1(q^2) + i \frac{[\gamma_\mu,\gamma_\nu]}2 q_\nu\; \frac{F_2(q^2)}{2 M_N} + (2 i\,M_N \gamma_5 q_\mu - \gamma_\mu \gamma_5 q^2)\;\frac{F_A(q^2)}{M_N^2} \right. \nonumber \\
&&\qquad\qquad\left.
         {} + \frac{[\gamma_\mu, \gamma_\nu]}2 q_\nu \gamma_5\;\frac{F_3(q^2)}{2 M_N} \right] u_N\,.
\end{eqnarray}
Here \(u_N\) is the standard Dirac spinor satisfying \(\slashed p u_N = M_N u_N\), but, in the absence of parity symmetry, it is not the wavefunction of the asymptotic state created by a generic interpolating field, \(\hat N\).  Instead, we generally have \(\langle\Omega|\hat N|N\rangle \propto e^{i\alpha_N\gamma_5}u_N\), where \(\alpha_N\) is a state dependent factor; for our standard choices for the interpolating operator \(\hat N\), CP symmetry of the theory implies \(\Im \alpha_N=0\), and PT symmetry implies \(\Re \alpha_N=0\).

The results presented here use the clover-on-HISQ formulation: clover valence quarks on HYP-smeared~\cite{PhysRevD.64.034504} HISQ ensembles generated by the MILC collaboration~\cite{Bazavov:2012xda}.  The clover coefficient is fixed at its tadpole-improved value \(c_{sw}=1/u_0^3\), where \(u_0\) is the fourth root of the plaquette on smeared lattices. Details of the ensembles are given in our previous publications~\cite{PhysRevD.103.114507}. 


\section{QCD Topological Term}

The coefficient, \(\Theta\), of the QCD topological charge is constrained to a tiny value since otherwise the neutron electric dipole moment (nEDM) would already have been observed~\cite{PhysRevLett.124.081803,Baker:2006ts,Graner:2016ses}. But the precise relation between the nEDM and \(\Theta\) has been difficult to find on the lattice since the ultraviolet fluctuations of the topological charge density are large.  Gradient flow~\cite{Luescher2010} smoothens these ultraviolet fluctuations, and we use this technique here to study the effect of the \(\Theta\)-term.
%
\begin{figure}
  \begin{minipage}{0.99\hsize}
  \begin{center}
    \includegraphics[width=0.425\hsize]{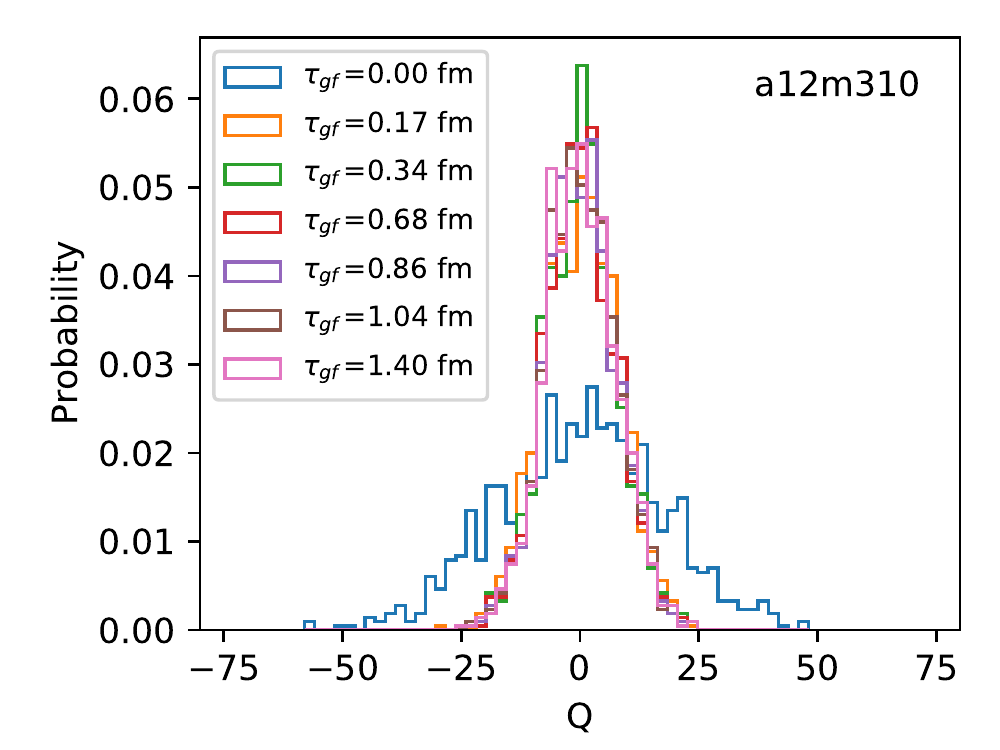}
    \includegraphics[width=0.425\hsize]{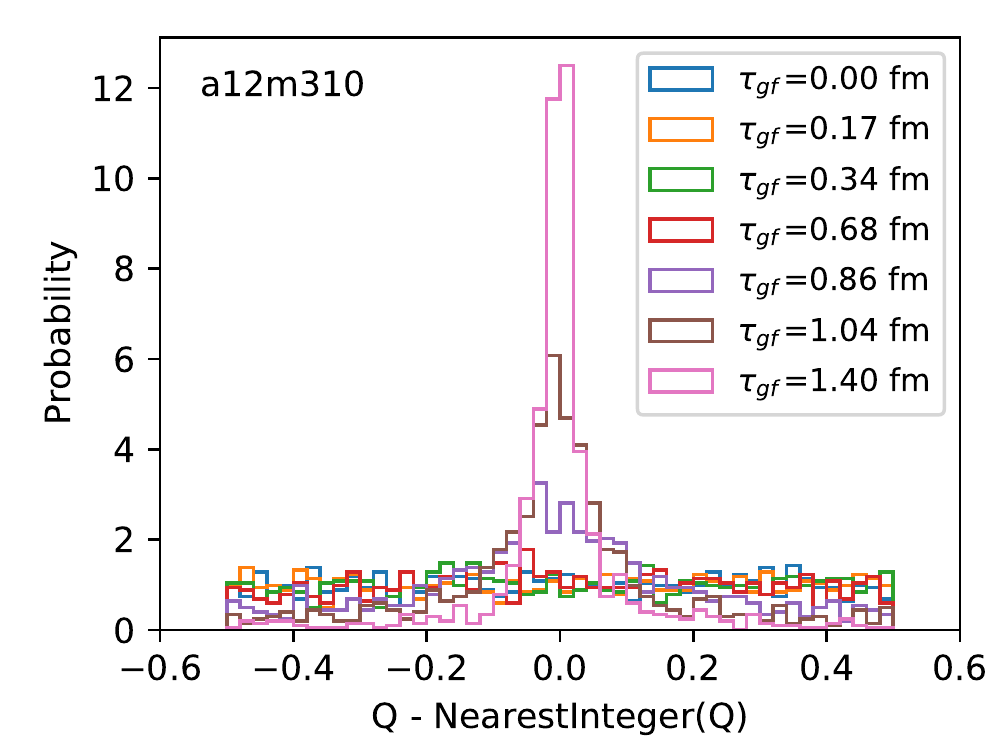}
  \end{center}
  \caption{The topological charge distribution (left) stabilizes with a very small amount of gradient flow, but the noninteger part (right) survives till much later.}
  \label{fig:Qdistribution}
\end{minipage}
\end{figure}
As shown in Fig.~\ref{fig:Qdistribution}, the overall topological charge distribution stabilizes even with a small amount of gradient flow.  The charge, however, does not become an integer until much later in the flow.  In all our calculations, we use the small-$\Theta$ expansion instead of weighting the path integral with a phase proportional to the topological charge. For this reason, we are not overly sensitive to the the topological charge taking on noninteger values.  To be conservative, we use a value of the flow-time, \(\tau_{\rm gf}\equiv\sqrt{8t_{\rm gf}}=0.68~\rm fm\) for the \(a\approx0.06\) and \(0.09~\rm fm\) ensembles, and \(\tau_{\rm gf}=0.86~\rm fm\) for the \(a\approx0.12~\rm fm\) ensembles.

We first check the efficacy of the gradient-flow in suppressing ultraviolet fluctuations by comparing the topological susceptibility against that expected from Chiral Perturbation Theory ($\chi$PT).  The Witten-Veneziano relation~\cite{Witten:1979vv,Veneziano:1979ec}, modified by SU(3) breaking~\cite{Bernard:2017npd}, is:
\begin{equation}
\chi_Q^{\rm quench.} \approx \frac{F_\pi^2(M_{\eta'}^2-M_\eta^2)}6
                    \left(1 + 2 \frac{M_\eta^2-M_K^2}{M_{\eta'}^2-M_\eta^2}\right) \,.
\end{equation}
In the quenched limit this gives the topological susceptibility to be \(\chi^{\rm quench.}_Q \approx (179\ \rm MeV)^4\). With dynamical quarks~\cite{PhysRevD.103.114507}, the prediction is
\begin{equation}
\frac{1}{\chi_Q} \approx \frac1{\chi_Q^{\rm quench.}} +
              \frac4{M_\pi^2F_\pi^2}\left(1-\frac{M_\pi^2}{3M_\eta^2}\right)^{-1} \,,
\end{equation}
which gives \(\chi_Q\approx (79\ {\rm MeV})^4\).


\begin{figure}
\begin{minipage}{0.99\hsize}
  \begin{center}
    \includegraphics[width=0.425\hsize]{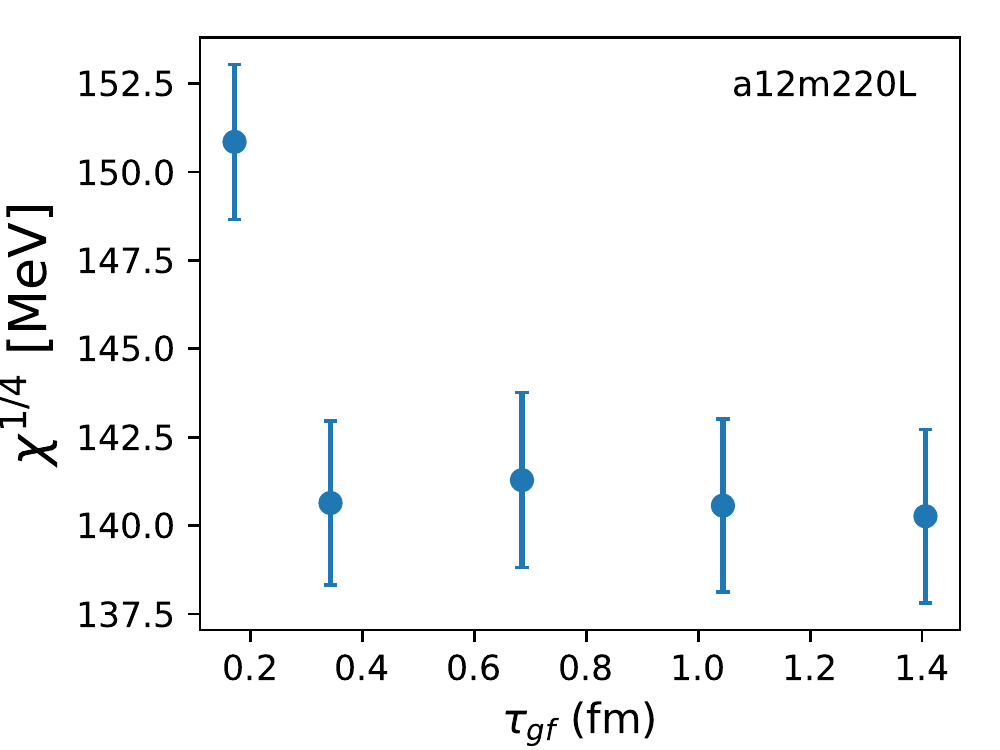}
    \includegraphics[width=0.425\hsize]{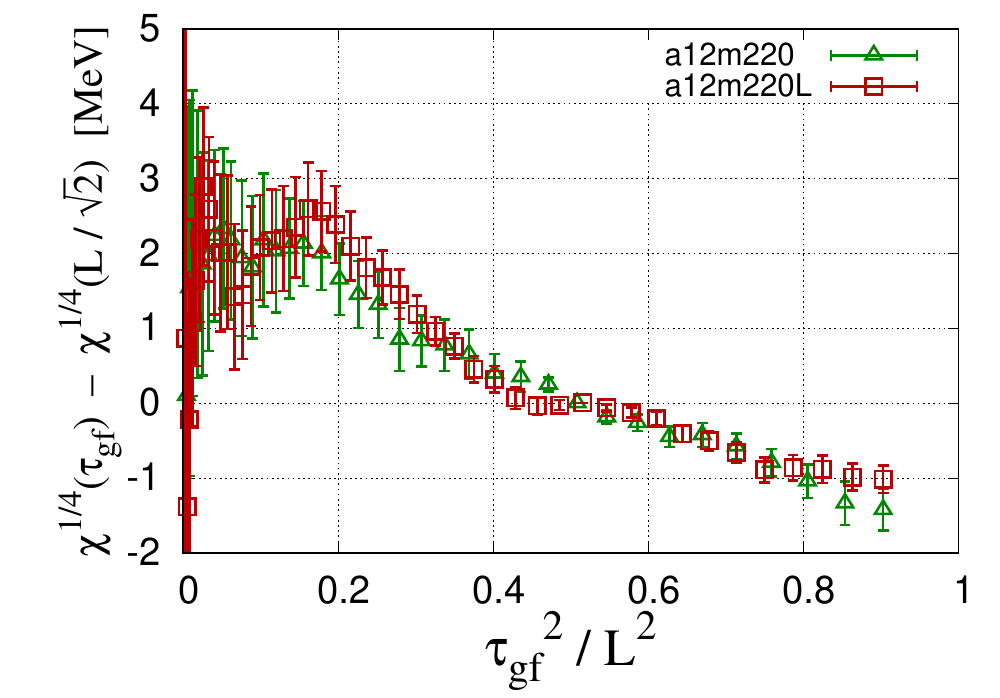}
  \end{center}
  \caption{The topological susceptibility becomes almost a constant after a small amount of gradient flow (left). Long flow-time behavior (right) is seen to be a finite volume effect.}
  \label{fig:topsusc}
\end{minipage}
\end{figure}
\begin{figure}[b]
\begin{minipage}{0.99\hsize}
  \begin{center}
    \includegraphics[width=0.425\hsize]{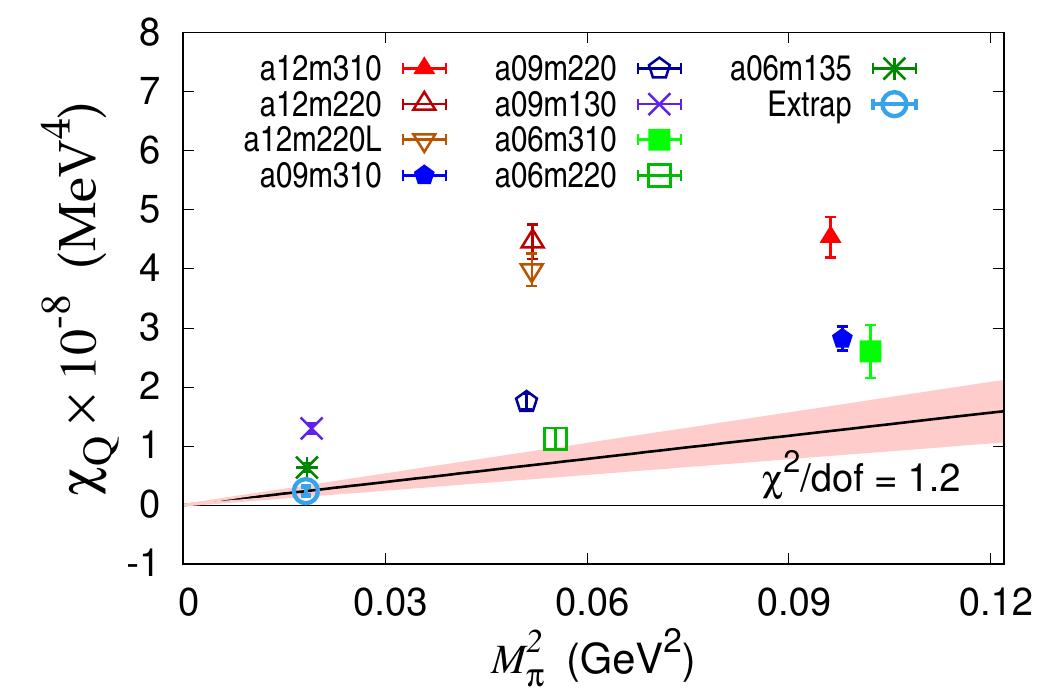}
    \includegraphics[width=0.425\hsize]{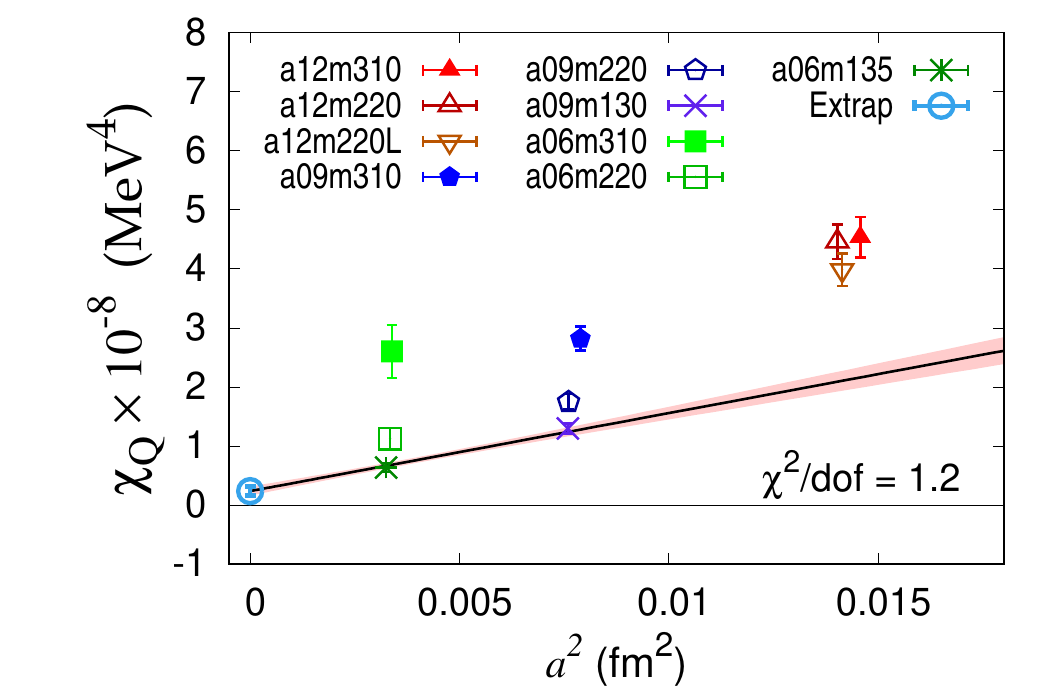}
  \end{center}
\caption{Extrapolation of the topological susceptibility,
ignoring the \(a12\) ensembles.}
\label{fig:Qchiralcont}
\end{minipage}
\end{figure}
To compare these with lattice results, we note that the topological charge in gradient-flow scheme does not need renormalization.  So, one expects that the lattice data for topological susceptibility should be independent of flow-time after discretization effects become negligible.  As shown in Fig.~\ref{fig:topsusc}, this result is almost correct. By comparing data on two different volumes, we show that the small downward trend at large flow-times is a finite volume effect.


We extrapolate the results to \(a=0\) and \(M_\pi=140~\rm MeV\) using
\(\chi_Q \approx c_1 a^2 + c_2 M_\pi^2 + c_3 a^2 M_\pi^2\).
The data at the coarsest lattices, \(a\approx 0.12\ \rm fm\) ensembles, do not fit, and for now we ignore it. The data at \(a\approx0.06\ \rm fm\), \(M_\pi\approx310\ \rm MeV\) may suffer from long correlation times (frozen topological charge).  The fit in Fig.~\ref{fig:Qchiralcont} including this a06m310 ensemble is, however, reasonable, and our final result with systematics included is \(\chi_Q=(66(9)(4)\ {\rm MeV})^4\), in good agreement with the $\chi$PT prediction.

\begin{figure}
\begin{minipage}{0.99\hsize}
  \begin{center}
    \includegraphics[width=0.425\hsize]{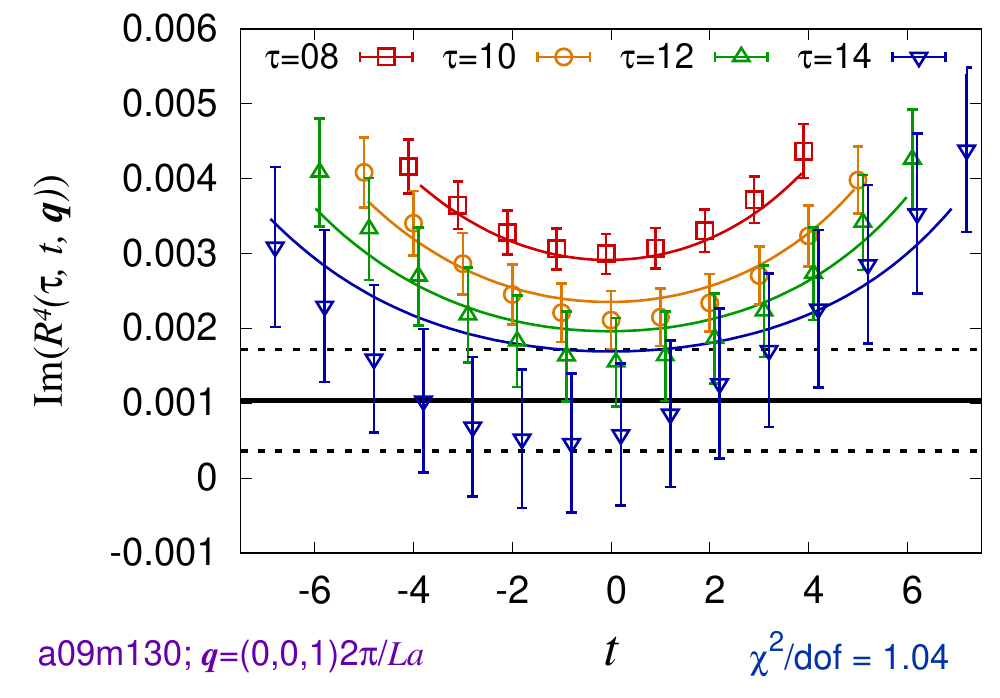}
    \includegraphics[width=0.425\hsize]{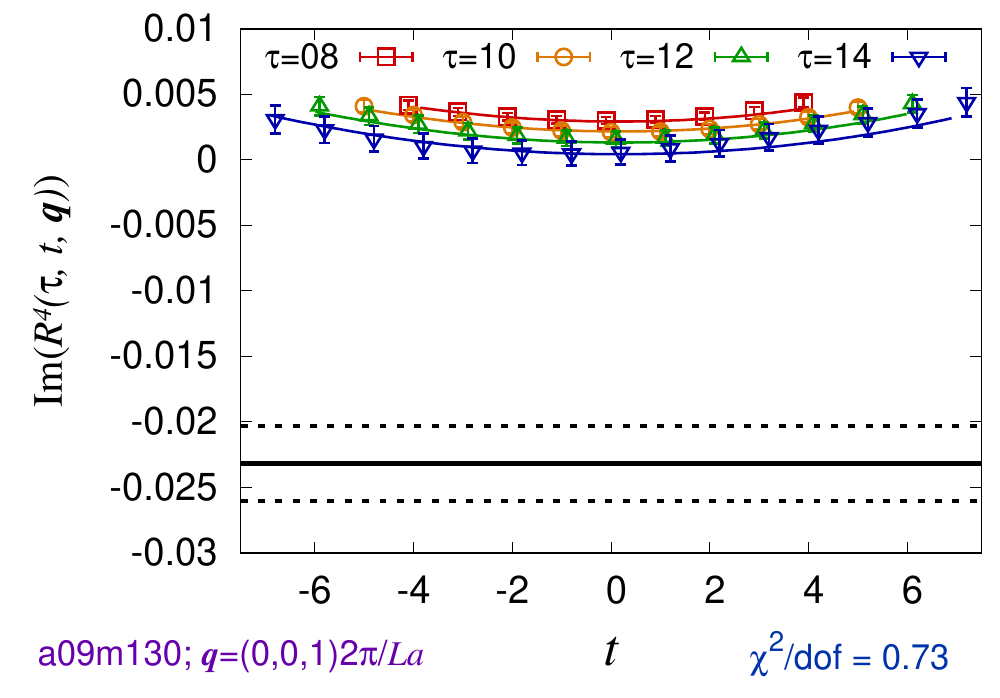}
  \end{center}
  \caption{Fits to remove ESC in the three-point function of \(\Im V_4\) in the presence of \(\Theta\).
  The left panel shows the standard analysis where the spectrum is obtained from fits to the two-point function, whereas the right panel assumes the dominant contribution is due to an \(N\pi\) intermediate state.}
  \label{fig:ESC}
\end{minipage}
\end{figure}
\begin{figure}[b]
  \begin{center}
    \includegraphics[width=0.425\hsize]{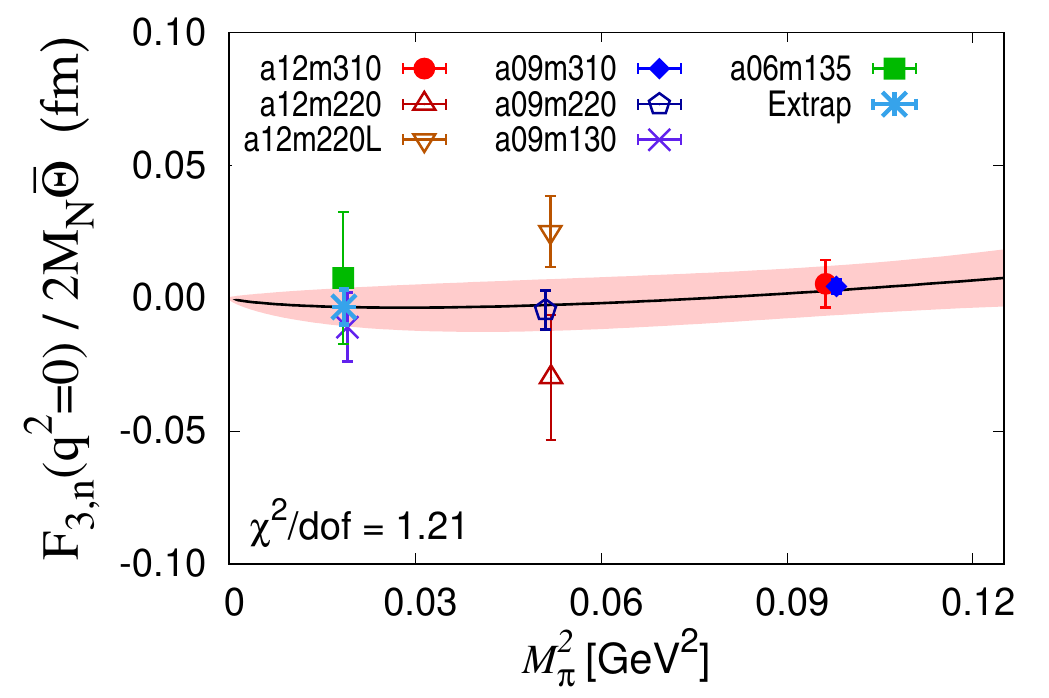}
    \includegraphics[width=0.425\hsize]{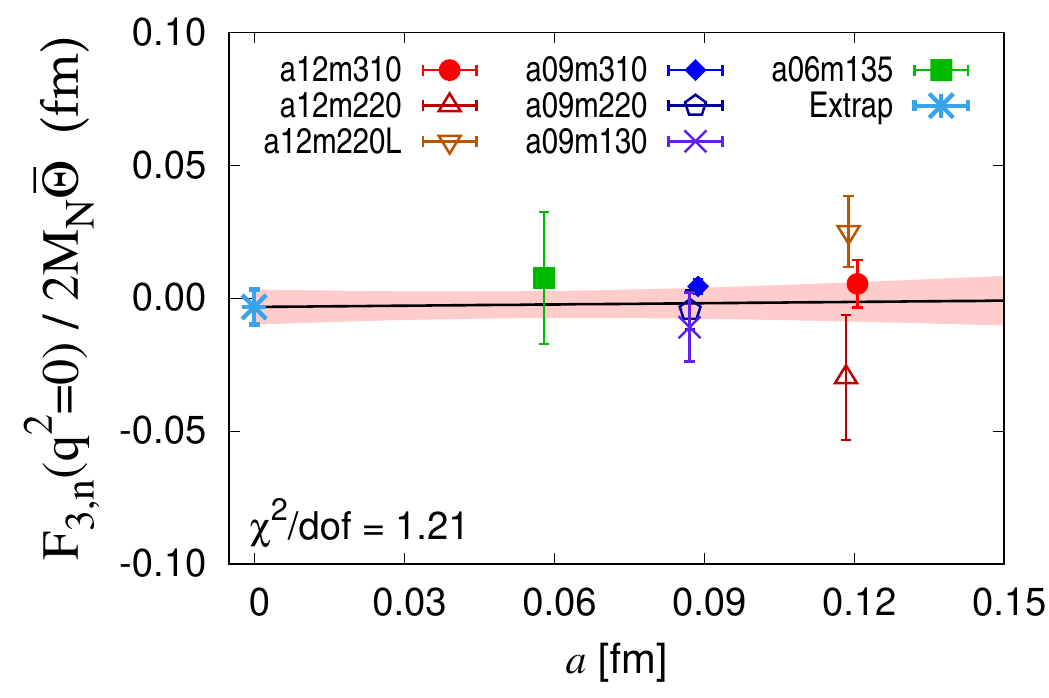}
  \end{center}
  \caption{Simultaneous chiral and continuum fit plotted versus $M_\pi^2$  (left) and $a$ (right) to get the nEDM due to the topological term. These data were obtained using the standard analysis to remove excited-state contamination.}
  \label{fig:CC}
\end{figure}

One of the major problems recently discovered~\cite{PhysRevLett.124.072002} about lattice calculations of baryon matrix elements is that the baryon sources create low-mass multihadron states.  Traditionally, the spectrum of these excited states were evaluated using multi-exponential fits to baryon two-point functions, and then used in fits to the three-point functions.  What we now  realize is that exponential fits are not very sensitive to low-lying excited states since
            \begin{equation}A + B e^{-\Delta t}\approx (A+B) - (B \Delta) t\qquad {\rm for}\qquad t \ll \Delta^{-1}\,,\end{equation}
            and it is difficult to obtain the required value \(A\) at \(t\to\infty\) using data at only moderate \(t\).
As a result, with finite precision data, the results depend greatly on the priors one puts on the excited state spectrum.  Thus, as shown in Fig.~\ref{fig:ESC}, the value of the ground-state matrix element depends strongly on whether one assumes an \(N\pi\) excited state makes a contribution, as expected by $\chi$PT, or whether the fits to the two-point function gives all the states that contribute significantly.


The electric dipole moment is obtained from the value of the form factor at \(Q^2=0\). Chiral perturbation theory provides guidance for \(Q^2\) fits. With our data, however, linear fits or fits without constraining the coefficient of the chiral logarithm makes only a small difference. 
%
%
%
To obtain the central results, we carry out a chiral fit based on $\chi$PT and assume a linear dependence on \(a\); this is shown in Fig.~\ref{fig:CC}.  The final results are 
\begin{eqnarray}
        d_n = -0.003(7)(20)\ \,\overline\Theta e\cdot\rm fm && d_p = 0.024(10)(30)\ \,\overline\Theta e\cdot\rm fm\qquad\rm\ (Standard\ analysis)\\
        d_n = -0.028(18)(54)\,\overline\Theta e\cdot\rm fm && d_p = 0.069(25)(120) \,\overline\Theta e\cdot\rm fm\qquad\rm(Assuming\ N\pi\ state)
\end{eqnarray}
where the second errors are estimates of the systematics other than due to the excited-state spectrum.

\section{Weinberg Three-Gluon Operator}

\begin{figure}
\begin{minipage}{0.99\hsize}
  \begin{center}
    \includegraphics[width=0.45\hsize]{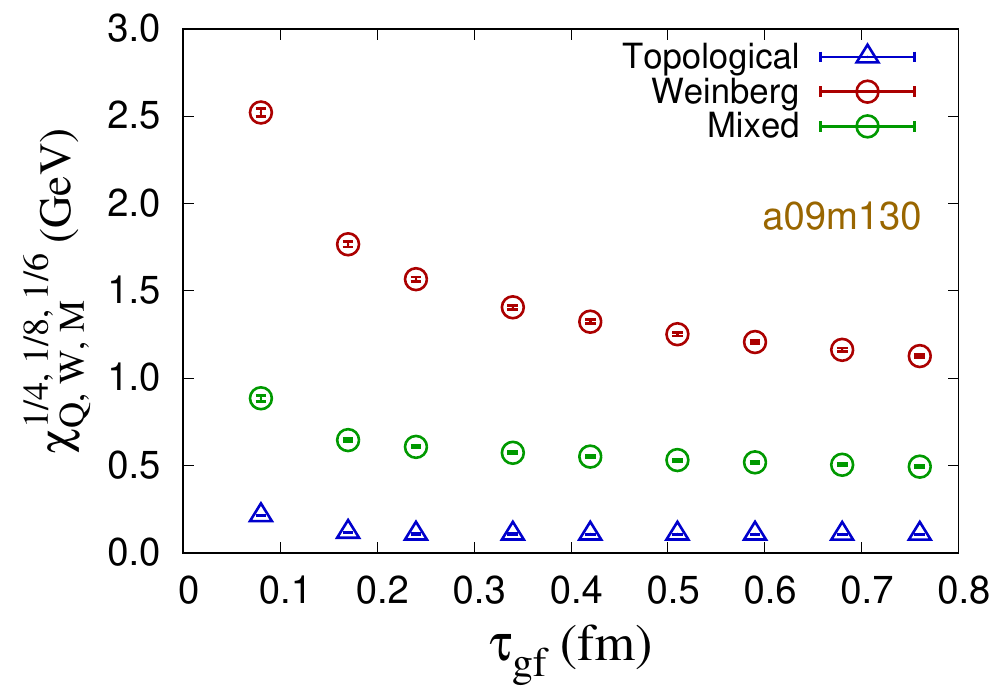}
    \vspace{-4ex}
  \end{center}
  \caption{The \(\sqrt[4]{\chi_Q}\) (topological, blue points at the bottom), \(\sqrt[8]{\chi_W}\) (Weinberg, red points on the top), and \(\sqrt[6]{\chi_M}\) (mixed, green points in the middle) susceptibilities for the a09m130 ensemble.}
  \label{fig:Wsusc}
  \end{minipage}
\end{figure}
\begin{figure}[b]
  \begin{minipage}{0.99\hsize}
  \begin{center}
    \includegraphics[width=0.425\hsize]{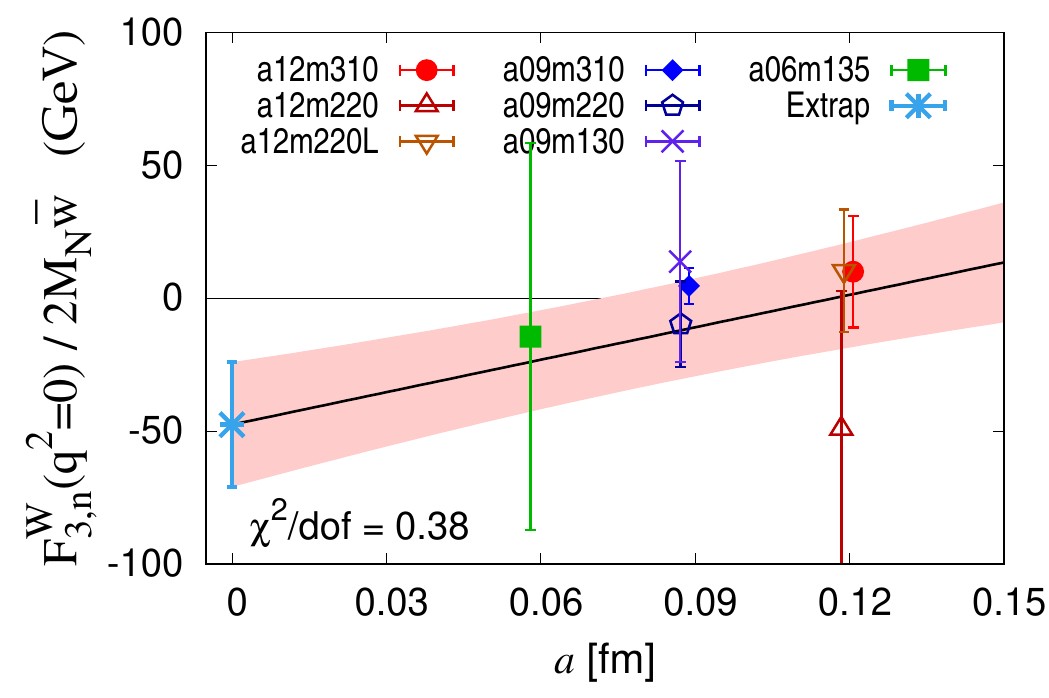}
    \includegraphics[width=0.425\hsize]{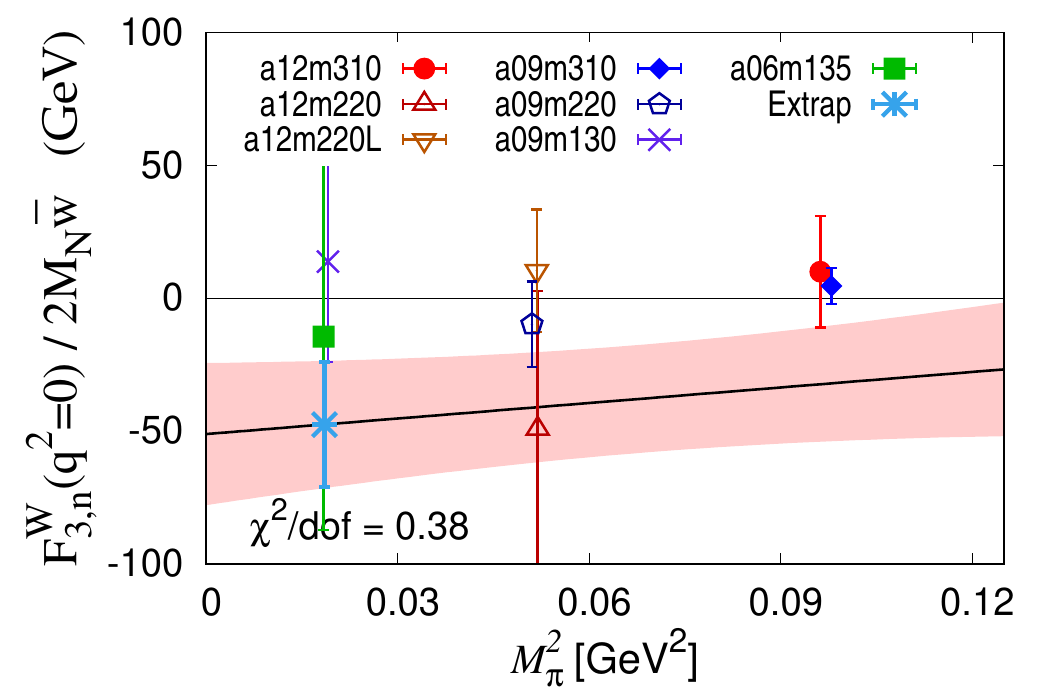}
  \end{center}
  \caption{The simultaneous chiral (right) continuum (left) extrapolation of nEDM due to Weinberg operator in the gradient-flow scheme at \(\tau_{\rm gf}\approx 0.34\)~fm.}
  \label{fig:WF3}
\end{minipage}
\end{figure}

The Weinberg three-gluon operator mixes with other operators of the same dimension, and with lower dimension operators like the topological term. The latter mixing is especially problematic since this diverges as we take the continuum limit.  In the gradient-flow scheme~\cite{Luescher2010}, however, the flow-time acts as a gauge- and Lorentz-symmetric hard ultraviolet cutoff, even as the chiral and rotational symmetric breaking lattice artifacts vanish in the continuum limit \(a\to0\). Thus, in this scheme, the matrix elements of the Weinberg operator have a finite continuum limit, but do have a \(O(1/t_{\rm gf})\) mixing with the lower dimensional topological charge~\cite{Rizik:2020naq}, a \(\log t_{\rm gf}\) mixing with operators of the same dimension, and an \(O(t_{\rm gf})\) mixing with higher dimension operators.  As a result, in contrast with the almost flow-time independent topological susceptibility, the Weinberg and mixed susceptibilities have a strong dependence on the flow time, as shown in Fig.~\ref{fig:Wsusc}. In addition, the \(F_3\) calculated from the matrix element of the product of the  vector current and the Weinberg operator, needs the subtraction of a \(\log t_{\rm gf}\) dependent contribution from the quark-EDM operator.


In Fig.~\ref{fig:WF3}, we present the results in the continuum-extrapolated gradient-flow scheme at a fixed \(\tau_{\rm gf}\approx 0.34\)~fm.  
Note that the results are scheme and flow-time dependent. To convert these to a convenient scheme like $\overline{\rm MS}$ requires a perturbative calculation of the mixing and renormalization constants.

\section{Quark Chromoelectric Dipole Moment}
\label{sec:qcEDM}

\begin{figure}[t]
  \begin{minipage}{0.99\hsize}
  \vspace*{\baselineskip}
  \begin{center}
    \includegraphics[width=0.45\hsize]{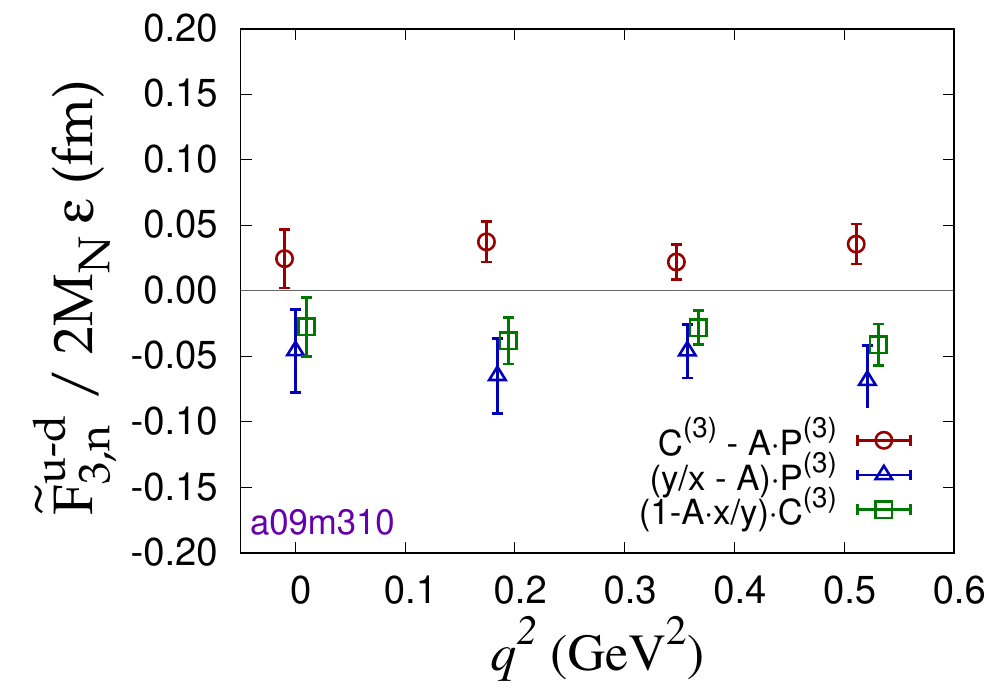}\qquad
    \includegraphics[width=0.45\hsize]{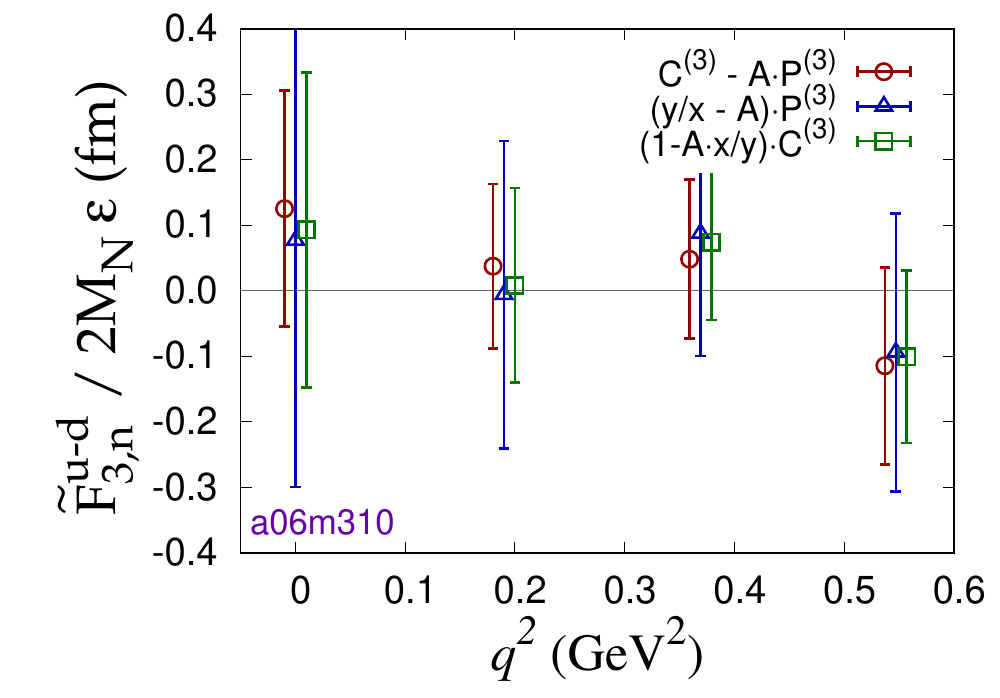}
  \end{center}
  \caption{Different methods discussed in Sec.~\ref{sec:qcEDM} for subtracting the power-law divergence in the cEDM operator on two example ensembles.  $q^2=0$ results are obtained by a linear extrapolation from $q^2>0$ data.}
  \label{fig:xy}
  \end{minipage}
\end{figure}
Like the Weinberg operator, the chromo-electric dipole moment (cEDM) operator has power-divergent mixing with the pseudoscalar operator, which can be removed by gradient-flow techniques. In this work, we present an alternate analysis using only unflowed results.  To this end, we notice that the power-divergence can be subtracted to define an operator
\begin{equation}
\tilde C \equiv i\bar\psi \sigma^{\mu\nu}\gamma_5 G_{\mu\nu}T^a\psi
- \frac{iA}{a^2}\bar\psi\gamma_5 T^a\psi\,,
\end{equation}
which has only logarithmic mixing for on-shell zero four-momentum quantities. Here the coefficient \(A\) needs to be adjusted to cancel the ultraviolet divergence, and we fix it by requiring that \(\tilde C\) not create a single pion out of the vacuum, i.e., \(\langle \Omega|\tilde C|\pi(\vec p=0)\rangle=0\).

Furthermore, at zero four-momentum, any isovector \(P\) can be rotated away using the nonanomalous Ward identity!  For Wilson-clover fermions, however, the hard violation of chiral symmetry leaves behind an \(O(a)\) piece after this rotation:
\begin{equation}
  Z_A(1+b_Ama)\partial\cdot A+iaZ_Ac_A\partial^2 P + 2 miP-iaK\tilde C\sim0
  \label{eq:WI}
\end{equation}

\begin{figure}[t]
  \begin{minipage}{0.99\hsize}
  \begin{center}
    \includegraphics[width=0.425\hsize]{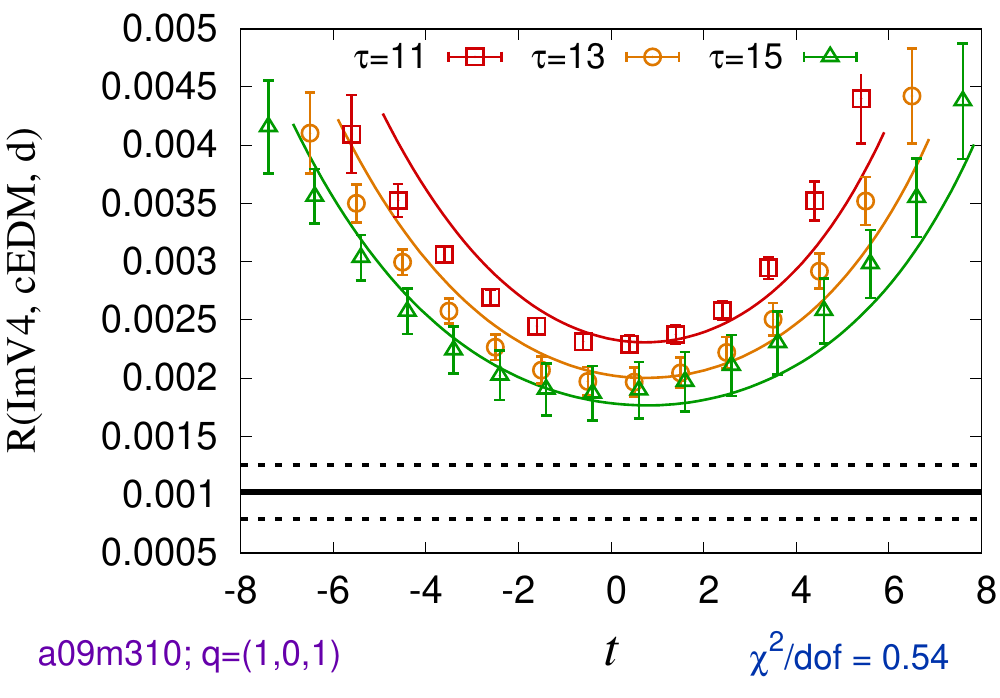}
    \includegraphics[width=0.425\hsize]{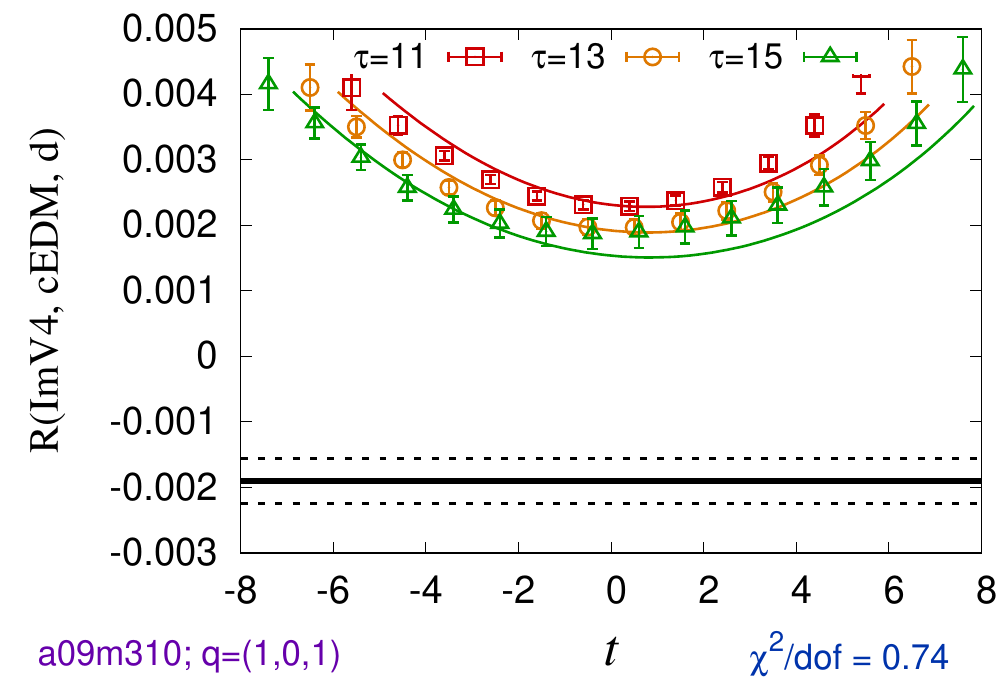}
  \end{center}
  \caption{Example of fits to remove ESC in the three-point function of the imaginary part \(V_4\) in the presence of a cEDM on the \(d\) quark. The rest is similar to Fig.~\ref{fig:ESC}.}
  \label{fig:Npicomp}
\end{minipage}
\end{figure}
The constant \(K\) that appears in Eq.~\eqref{eq:WI} can be evaluated by implementing the Ward identity on any state. To obtain good signal, we choose the state \(|\pi\rangle\) created by a smeared pseudoscalar interpolating field.
\begin{equation}
  \frac{ia\Delta_4\langle\pi|A_4\rangle-\bar c_A a^2\Delta_4^2\langle\pi|P\rangle+\bar x a^2\langle\pi|C\rangle}{\langle\pi|P\rangle} \approx \bar y + O(a^2)\,
  \label{eq:pionWI}
\end{equation}
where \(A_4 (x) =  \bar{\psi}  (x) T^3 \gamma_4 \gamma_5 \psi  (x) \), \(C \equiv i\,  \bar\psi \sigma^{\mu\nu} \gamma_5 G_{\mu\nu}  T^3\psi\), \(P \equiv \bar\psi i\gamma_5  T^3 \psi\), \(x \equiv K\), \(y \equiv 2 m a + A\), and the bar denotes division by \(1 + b_Ama\). We can determine \(\bar c_A\), \(\bar x\) and \(\bar y\) by fitting the LHS to a constant; in practice, we first determine \(\bar c_A\) to make the LHS a constant over a large range in Euclidean time, and then determine \(\bar x\) and \(\bar y\) to extend the region holding \(\bar c_A\) fixed.
In terms of these, the three isovector operators
\(\left(aC - A \frac Pa\right)\), \(\left(\frac{\bar y}{\bar x} - A\right) \frac Pa\), and \(\left(1 - \frac {A\bar x}{\bar y}\right) a C\) are related by the Ward identity, and describe the same physics!  Note that the coefficient \(\bar x\) is zero if all \(O(a)\) chiral symmetry breaking is removed by nonperturbatively tuning the coefficient \(c_{SW}\) of the clover term.  But, as is clear from the equations, the physical effects of the isovector operator \(P\) are enhanced by one inverse power of \(ma\), so a small mistuning gives a large contribution.  By the same token, any \(O(a^2)\) effect in the determination of \(\bar x/\bar y\) from the Ward identity have a large effect on the determination of \(F_3\).  As a result, the different determinations of the contribution of the cEDM operator do not agree, as shown in Fig.~\ref{fig:xy}; leading to a large systematic error in the determination of \(F_3\). \looseness-1

\begin{figure}[b]
  \begin{center}
    \includegraphics[width=0.4\hsize]{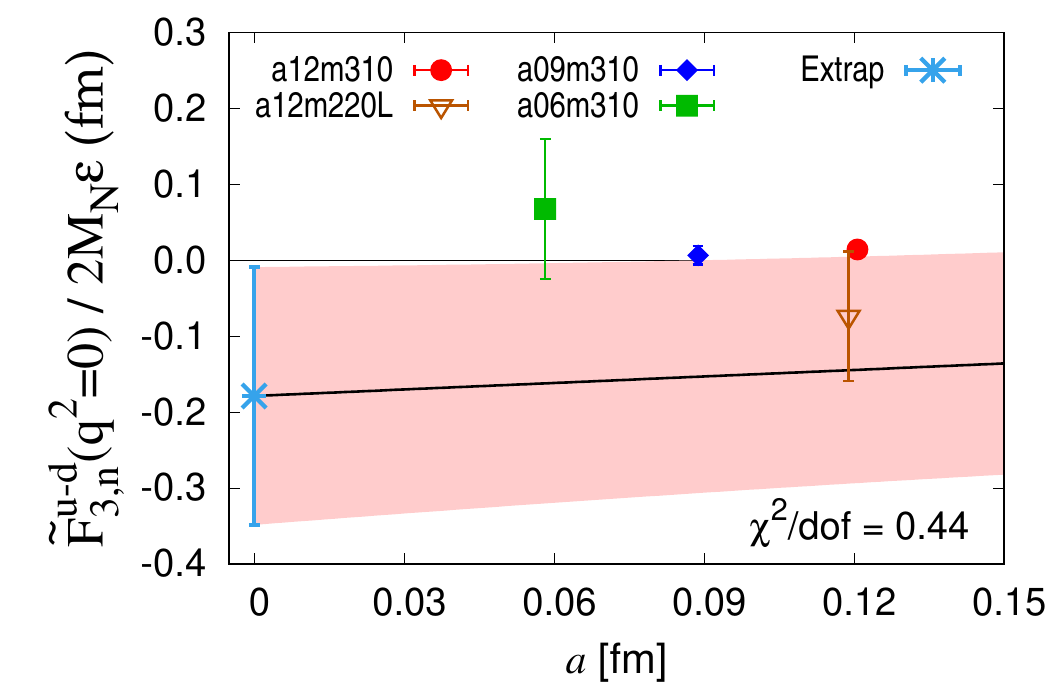}\qquad
    \includegraphics[width=0.4\hsize]{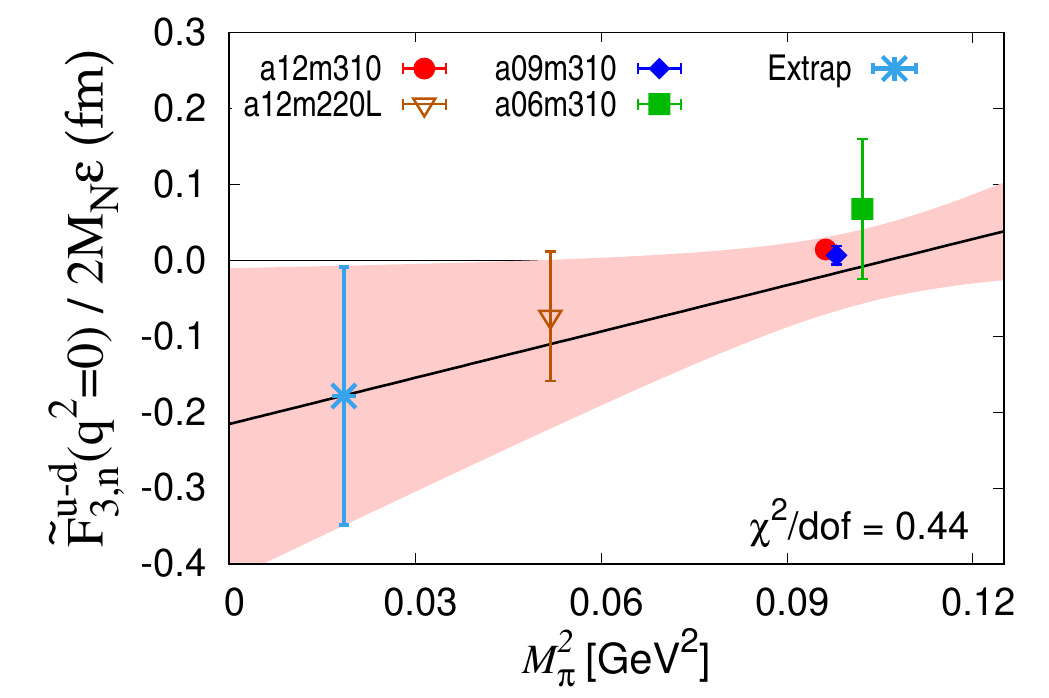}
  \end{center}
  \vspace{-3ex}
  \caption{Chiral-continuum extrapolation of nEDM due to cEDM using the fit ansatz: $d_N = c_1 + c_2 M_\pi^2 + c_3 a$ giving $-0.18(17)$ at the physical point. A fit to \(a^2\) has almost same quality and the same extrapolated value. On each ensemble, the excited state contamination is removed using the standard analysis based on the spectrum from the nucleon two-point function.}
  \label{fig:cedmCC}
\end{figure}

In addition to this uncertainty, the assumed spectrum of intermediate states that contribute significantly to the three-point function also makes a large difference in the determination. Fig.~\ref{fig:Npicomp} illustrates this uncertainty.
%
%
Finally, we show a preliminary chiral-continuum extrapolation of the results in Fig.~\ref{fig:cedmCC}.  The errors indicated here are only statistical.


\section{Conclusions}
Our calculations show that there is no reliable determination of nEDM from the topological term yet.  We find that gradient-flow is a good approach to study gluonic quantities. A very important systematics one needs to tackle is understanding the spectrum of low-lying excited states that give significant contribution to the three-point function. Finally, we showed that, in principle, the power-divergence in isovector cEDM does not pose a problem, but these are not yet under control in current calculations.

We thank the MILC collaboration for the HISQ ensembles~\cite{Bazavov:2012xda}. The calculations used the Chroma software suite~\cite{Edwards:2004sx}. This research used resources at NERSC, ORNL, the USQCD Collaboration and Institutional Computing at Los Alamos National Laboratory. 
Parts of the work were supported by the U.S. DOE Offices of Science, HEP and NP, through LANL, and by the LANL LDRD program.

\bibliographystyle{JHEP}
\let\oldbibitem\bibitem
\def\bibitem#1\emph#2,{\oldbibitem#1}
\let\oldthebibliography\thebibliography
\renewcommand\thebibliography[1]{\oldthebibliography{#1}%
                                 \itemsep0pt\parskip0pt\relax}
\bibliography{refs}

\providecommand{\href}[2]{#2}\begingroup\raggedright\begin{thebibliography}{10}

\bibitem{PhysRevLett.10.531}
N.~Cabibbo, \emph{Unitary symmetry and leptonic decays},
  \href{https://doi.org/10.1103/PhysRevLett.10.531}{\emph{Phys. Rev. Lett.}
  {\bfseries 10} (1963) 531}.

\bibitem{10.1143/PTP.49.652}
M.~Kobayashi and T.~Maskawa, \emph{{CP}-violation in the renormalizable theory
  of weak interaction},
  \href{https://doi.org/10.1143/PTP.49.652}{\emph{Progress of Theoretical
  Physics} {\bfseries 49} (1973) 652}.

\bibitem{10.1143/PTP.28.870}
Z.~Maki, M.~Nakagawa and S.~Sakata, \emph{{Remarks on the Unified Model of
  Elementary Particles}},
  \href{https://doi.org/10.1143/PTP.28.870}{\emph{Progress of Theoretical
  Physics} {\bfseries 28} (1962) 870}.

\bibitem{Pontecorvo:1957qd}
B.~Pontecorvo, \emph{Inverse beta processes and nonconservation of lepton
  charge}, {\emph{Zh. Eksp. Teor. Fiz.} {\bfseries 34} (1957) 247}.

\bibitem{Baker:2006ts}
C.~A. Baker et~al., \emph{An improved experimental limit on the electric dipole
  moment of the neutron},
  \href{https://doi.org/10.1103/PhysRevLett.97.131801}{\emph{Phys. Rev. Lett.}
  {\bfseries 97} (2006) 131801}.

\bibitem{Graner:2016ses}
B.~Graner, Y.~Chen, E.~G. Lindahl and B.~R. Heckel, \emph{Reduced limit on the
  permanent electric dipole moment of {Hg199}},
  \href{https://doi.org/10.1103/PhysRevLett.116.161601}{\emph{Phys. Rev. Lett.}
  {\bfseries 116} (2016) 161601}.

\bibitem{Bennett:2003bz}
{\scshape {WMAP}} collaboration, C.~L. Bennett et~al., \emph{First year
  {W}ilkinson {M}icrowave {A}nisotropy {P}robe ({WMAP}) observations:
  Preliminary maps and basic results},
  \href{https://doi.org/10.1086/377253}{\emph{Astrophys. J. Suppl.} {\bfseries
  148} (2003) 1}.

\bibitem{Sakharov:1967dj}
A.~D. Sakharov, \emph{Violation of {CP} invariance, {C} asymmetry, and baryon
  asymmetry of the universe},
  \href{https://doi.org/10.1070/PU1991v034n05ABEH002497}{\emph{Pisma Zh. Eksp.
  Teor. Fiz.} {\bfseries 5} (1967) 32}.

\bibitem{Dolgov:1991fr}
A.~D. Dolgov, \emph{Non{GUT} baryogenesis},
  \href{https://doi.org/10.1016/0370-1573(92)90107-B}{\emph{Phys. Rept.}
  {\bfseries 222} (1992) 309}.

\bibitem{Coppi:2004za}
P.~Coppi, \emph{How do we know antimater is absent?}, {\emph{eConf} {\bfseries
  C040802} (2004) L017}.

\bibitem{Pospelov:2005pr}
M.~Pospelov and A.~Ritz, \emph{Electric dipole moments as probes of new
  physics}, \href{https://doi.org/10.1016/j.aop.2005.04.002}{\emph{Annals
  Phys.} {\bfseries 318} (2005) 119}.

\bibitem{PhysRevD.64.034504}
A.~Hasenfratz and F.~Knechtli, \emph{Flavor symmetry and the static potential
  with hypercubic blocking},
  \href{https://doi.org/10.1103/PhysRevD.64.034504}{\emph{Phys. Rev. D}
  {\bfseries 64} (2001) 034504}.

\bibitem{Bazavov:2012xda}
{\scshape {MILC}} collaboration, A.~Bazavov et~al., \emph{Lattice {QCD}
  ensembles with four flavors of highly improved staggered quarks},
  \href{https://doi.org/10.1103/PhysRevD.87.054505}{\emph{Phys. Rev.}
  {\bfseries D87} (2013) 054505}.

\bibitem{PhysRevD.103.114507}
T.~Bhattacharya, V.~Cirigliano, R.~Gupta, E.~Mereghetti and B.~Yoon,
  \emph{Contribution of the {QCD} {$\mathrm{\ensuremath{\Theta}}$}-term to the
  nucleon electric dipole moment},
  \href{https://doi.org/10.1103/PhysRevD.103.114507}{\emph{Phys. Rev. D}
  {\bfseries 103} (2021) 114507}.

\bibitem{PhysRevLett.124.081803}
C.~Abel, S.~Afach, N.~J. Ayres, C.~A. Baker, G.~Ban, G.~Bison et~al.,
  \emph{Measurement of the permanent electric dipole moment of the neutron},
  \href{https://doi.org/10.1103/PhysRevLett.124.081803}{\emph{Phys. Rev. Lett.}
  {\bfseries 124} (2020) 081803}.

\bibitem{Luescher2010}
M.~L{\"u}scher, \emph{Properties and uses of the {W}ilson flow in lattice
  {QCD}}, \href{https://doi.org/10.1007/JHEP08(2010)071}{\emph{Journal of High
  Energy Physics} {\bfseries 2010} (2010) 71}.

\bibitem{Witten:1979vv}
E.~Witten, \emph{Current algebra theorems for the {U(1)} {G}oldstone boson},
  \href{https://doi.org/10.1016/0550-3213(79)90031-2}{\emph{Nucl. Phys. B}
  {\bfseries 156} (1979) 269}.

\bibitem{Veneziano:1979ec}
G.~Veneziano, \emph{{U(1)} without instantons},
  \href{https://doi.org/10.1016/0550-3213(79)90332-8}{\emph{Nucl. Phys. B}
  {\bfseries 159} (1979) 213}.

\bibitem{Bernard:2017npd}
{\scshape MILC} collaboration, C.~Bernard and D.~Toussaint, \emph{Effects of
  nonequilibrated topological charge distributions on pseudoscalar meson masses
  and decay constants},
  \href{https://doi.org/10.1103/PhysRevD.97.074502}{\emph{Phys. Rev. D}
  {\bfseries 97} (2018) 074502}.

\bibitem{PhysRevLett.124.072002}
Y.-C. Jang, R.~Gupta, B.~Yoon and T.~Bhattacharya, \emph{Axial vector form
  factors from lattice {QCD} that satisfy the {PCAC} relation},
  \href{https://doi.org/10.1103/PhysRevLett.124.072002}{\emph{Phys. Rev. Lett.}
  {\bfseries 124} (2020) 072002}.

\bibitem{Rizik:2020naq}
{\scshape SymLat} collaboration, M.~D. Rizik, C.~J. Monahan and A.~Shindler,
  \emph{Short flow-time coefficients of {$CP$}-violating operators},
  \href{https://doi.org/10.1103/PhysRevD.102.034509}{\emph{Phys.\ Rev.\ D}
  {\bfseries 102} (2020) 034509}.

\bibitem{Edwards:2004sx}
{\scshape {SciDAC}, {LHPC}, {UKQCD}} collaboration, R.~G. Edwards and B.~Joo,
  \emph{The {C}hroma software system for lattice {QCD}},
  \href{https://doi.org/10.1016/j.nuclphysbps.2004.11.254}{\emph{Nucl. Phys.
  Proc. Suppl.} {\bfseries 140} (2005) 832}.

\end{thebibliography}\endgroup
\end{document}